# Al substitution in MgB$_2$ crystals: influence on superconducting and structural properties

J. Karpinski, N. D. Zhigadlo, G. Schuck, S. M. Kazakov, B. Batlogg, K. Rogacki[1], R. Puzniak[2], J. Jun, E. Müller, P. Wägli,

Laboratory for Solid State Physics, ETH, 8093 Zürich, Switzerland

[1]also at Institute of Low Temperature and Structure Research, Polish Academy of Sciences, 50-950 Wroclaw, Poland

[2] also at Institute of Physics, Polish Academy of Sciences, PL 02-668 Warsaw, Poland

R. Gonnelli, D. Daghero, G. A. Ummarino, Dipartimento di Fisica, Politecnico di Torino, 10129 Torino, Italy

V. A. Stepanov, P. N. Lebedev Physical Institute, Russian Academy of Sciences, 119991 Moscow, Russia

Single crystals of Mg$_{1-x}$Al$_x$B$_2$ have been grown at a pressure of 30 kbar using the cubic anvil technique. Precipitation free crystals with $x < 0.1$ were obtained as a result of optimization of already developed MgB$_2$ crystal growth procedure. Systematic decrease of the $c$-axis lattice constant with increasing Al content, when the $a$-axis lattice constant is practically unchanged, was observed. Variation of the critical temperature on Al content in Mg$_{1-x}$Al$_x$B$_2$ crystals was found to be slightly different than that one observed for polycrystalline samples since, even a very small substitution of 1-2% of Al leads to the decrease of $T_c$ by about 2-3 K. X-ray and high resolution transmission electron microscopy investigations indicate on the appearance of second precipitation phase in the crystals with $x > 0.1$. This is in a form of non-superconducting MgAlB$_4$ domains in the structure of superconducting Mg$_{1-x}$Al$_x$B$_2$ matrix. Resistivity and magnetic investigations show the slight increase of the upper critical field, $H_{c2}$, for $H//c$ for the samples with small $x$, significant reduction of the $H_{c2}$ anisotropy at lower temperatures, and decrease of the residual resistance ratio value for Al substituted samples as compared to those of unsubstituted crystals. Superconducting gaps variation as a function of Al content, investigated with point contact spectroscopy for the series of the crystals with $T_c$ in the range from 20 to 37 K, does not indicate on the merging of the gaps with decreasing $T_c$ down to 20 K. It may be related to an appearance of the precipitation phase in the Mg$_{1-x}$Al$_x$B$_2$ structure.

# I. INTRODUCTION

MgB$_2$ is a two-gap superconductor with several anomalous properties originating from the existence of two separate sheets of the Fermi surface, one quasi 2D (σ band) and second quasi 3D (π band) [1, 2]. Without any scattering of electrons by phonons from one band to the other one, there would be two transition temperatures. In the case of weak, but finite, interband phonon scattering, the lower $T_c$ disappears and the temperature dependence of the lower gap becomes strongly non-BCS. This leads to high critical temperature of 39 K and temperature and field dependent anisotropy of MgB$_2$ [3]. The behavior of two gaps as a function of temperature and field have been studied intensively by point contact spectroscopy (PCS) [4, 5] and scanning tunneling spectroscopy (STS) [6 - 8].

In order to study the influence of intra- and interband scattering on the gap and superconducting properties the investigations of partially substituted MgB$_2$ crystals are extremely valuable. Theory predicts merging of two gaps with increased interband scattering caused by the impurities or substitutions [9, 10]. There are two kinds of substitutions, which were frequently studied in MgB$_2$: carbon substitution for boron, which influences mostly intraband scattering in the 2D σ band, and aluminum substitution for magnesium, which influences mostly intraband scattering in 3D π band. In both cases additional electrons are introduced in the structure, leading to a decrease of $T_c$. Gurevich [11] studied theoretically the influence of substitutions on anisotropy parameters and predicted decrease of the anisotropy with increasing substitution level. Bussmann-Holder and Bianconi [12] predicted that with Al substitution small π gap increases and large σ gap decreases, leading to merging of two gaps.

Up to now the substitutions in MgB$_2$ were investigated mainly on polycrystalline samples [13, 14]. Recently, several papers appeared on investigations of carbon and aluminum substituted MgB$_2$ crystals. Carbon substitution leads to increase of the upper critical fields $H_{c2}$ and to decrease of the $H_{c2}$ anisotropy γ [15]. Kang et al. [16] investigated the effect of Al substitution on the anisotropy by measuring the temperature dependence of the upper critical field from reversible magnetization of Mg$_{0.88}$Al$_{0.12}$B$_2$ single crystals with $T_c$ = 31 K. They found that, Al substitution increases $H_{c2}^{//c}$, but not $H_{c2}^{//ab}$ and decreases the anisotropy, making it practically temperature independent, which suggest increased interband scattering.

In this paper we present the crystal growth of Al substituted MgB$_2$ and show the results of systematic structure, magnetic, resistivity, and PCS studies performed on those crystals. Our preliminary data on superconducting and structural properties obtained on single crystals of Mg$_{1-x}$Al$_x$B$_2$ have been published recently [17] but the scope of the report was very strongly limited. Recent paper substantially extends the range of the investigations on the impact of Al substitution on crystallographic and superconducting properties of single crystals of $Mg_{1-x}Al_xB_2$. Carbon substitution results were reported in separate papers [18] and are not discussed in the present one.

Although the structure of $MgB_2$ is rather simple, the Al substitution chemistry is quite complicated. Structural investigations of Al substituted $MgB_2$ polycrystalline samples show the existence of a miscibility gap. For $Mg_{1-x}Al_xB_2$ polycrystalline samples of nominal compositions $0.1 < x < 0.4$, phase separation in two phases was observed [13]. These two phases have lattice parameters $c$ in between those of the samples with $x = 0.1$ and $0.4$. Our investigations on single crystals led to a different result: $Mg_{1-x}Al_xB_2$ crystals contain frequently a precipitation of the second phase. This is a non-superconducting $MgAlB_4$, which precipitates as Al rich domains. Small amount of this phase can be present in the crystals with any, even very small Al content and increases with increasing total Al content in the sample. However, we are going to show that, improved crystal growth procedure allows to obtain the single-phase crystals of $Mg_{1-x}Al_xB_2$ with Al content up to $x = 0.1$.

## II. EXPERIMENTAL

For the synthesis of Al substituted $MgB_2$ crystals we have applied a high-pressure growth method [19, 20]. Due to low solubility of $MgB_2$ in molten Mg at temperatures of 900 - 950°C used typically for synthesis of polycrystalline $MgB_2$ samples, it is difficult to grow crystals at normal pressure. In order to increase the solubility one has to increase the temperature significantly. Thus, one has to increase the pressure to stabilize the $MgB_2$ phase. At temperatures of 1860 - 1960°C used for crystal growth the Mg partial vapor pressure above liquid Mg reaches 100 bar. In order to keep the Mg partial pressure high we add excess of magnesium in the precursor. According to our phase diagram studies [20] high hydrostatic pressure above 20 kbar is necessary to stabilize the $MgB_2$ phase. A mixture of Mg and B is placed in a BN crucible in a pyrophyllite cube. The internal diameters of the crucibles are 6 or 8 mm, a length 7 mm. The heating element is a graphite tube. Six anvils generate pressure on the whole assembly. The typical process is as follows: (i) increasing of pressure up to 30 kbar, (ii) increasing of temperature up to 1860 - 1960°C in 1 h, (iii) dwelling for 0.5 - 1 h, (iv) decreasing the temperature and pressure in 1 h. At high temperature the reaction with the BN crucible is substantial. Nevertheless, BN is the only crucible material, which can be used for this process. Crystals of $MgB_2$ phase grew from a reaction in the ternary Mg-B-N system. This reaction is:

$4Mg + 8B + BN \rightarrow MgNB_9 + 3Mg \rightarrow MgB_2 + BN + 6B + 3Mg \rightarrow 4MgB_2 + BN$ (1)

Due to the reaction of the melt with the BN crucible new ternary nitride $MgNB_9$ [21] crystals grow, which in a later stage decompose forming seeds of $MgB_2$ crystals. As excess of Mg is available in the melt, further growth continues from the solution of boron (or $MgB_2$) in magnesium. Simultaneously transparent BN single crystals grow from the melt. After cooling the $MgB_2$ crystals are sticking together with BN crystals.

To substitute aluminum in $MgB_2$ crystals, two methods have been applied. To grow crystals with lower Al content up to about $x = 0.1$ part of Mg in the precursor was replaced by Al. Aluminum diffuses in the melt and partially substitutes magnesium. To grow crystals with larger Al content a disc made of Al was placed in the crucible together with Mg and B powder. In this way the $MgB_2$ crystals with Al content up to $x = 0.28$ were grown. The amount of Al determined by energy dispersive x-ray (EDX) in the crystals is lower than in the precursor and depends on the precursor composition and the growth temperature. Typical growth temperatures for Al substituted crystals were 1860 and 1960°C. Table I shows the nominal content of Al in the Mg-Al mixture used for crystal growth. With increasing growth temperature, $T_c$ of the crystals decreases and the Al content increased. This can be explained by higher solubility of Al in $Mg_{1-x}Al_xB_2$ at higher temperature. However, further increase of temperature up to 2000°C did not increase the Al content anymore. An increase of experimental time due to slow cooling led to a larger Al content and lower $T_c$. Using this method $Mg_{1-x}Al_xB_2$ single crystals of sizes up to 1 x 1 x 0.1 mm$^3$ have been grown (Fig. 1).

As it will be shown in the following chapters, single crystals of $Mg_{1-x}Al_xB_2$ with $x < 0.1$, grown with applied slow cooling procedure, were identified by us as single-phase material. It is not clear yet if there is any possibility to prevent precipitation of a second phase during crystal growth for higher Al content.

Obtained crystals have been investigated by means of x-ray single crystal diffractometer (Siemens P4 and Mar-300 Image Plate system), EDX, high resolution transmission electron microscopy (HRTEM), and PCS. Resistivity at various fields and currents was studied with Quantum Design Physical Properties and magnetization was studied with Magnetic Properties Measurement Systems. The diamagnetic dc susceptibility measurements of $Mg_{1-x}Al_xB_2$ crystals with various Al content were performed on home made SQUID magnetometer with Quantum Design sensor, The results of dc magnetization obtained with increasing temperature after zero field cooling are presented in Fig. 2.

## III. RESULTS AND DISCUSSION
### A. Structure analysis

The lattice parameters of Al-substituted crystals were determined by four-circle single crystal x-ray diffractometer Siemens P4. The same set of 40 reflections recorded in the wide range of $2\Theta$ angle (20 - 40 degree) was used to calculate the unit cell parameters. It was found that among the crystals with Al content $x \leq 0.1$ there are both multi- and single-phase samples, while all of the crystals with $x > 0.1$ show precipitation of an additional phase(s). In the case of multiphase samples, it was found that despite of dominant $Mg_{1-x}Al_xB_2$ phase, in the sample there is present one additional Al rich phase of $(Mg,Al)B_2$ and sometime an impurity phase as well.

The primary search was performed to determine the lattice parameters of single-phase sample or of the dominant phase, i.e., that one with lower Al content, if more than one phase exists. It is worth noting that with increasing Al content we observed a broadening of the reflections indicating increasing disorder. The measurements performed for the part of crystals with Mar-300 Image Plate system allowed us to determine independently lattice parameters of all of the phases in the multi-phase samples.

Figure 3a shows the $c$-axis parameter as a function of Al content determined with EDX. The relative variation of the $a$-axis lattice parameter is much smaller than that of the $c$-axis parameter, if any (Fig. 3b). The full circles show the lattice parameters, determined with Siemens P4 diffractometer, open up triangles show the lattice parameters of single-phase crystals and of the dominant $Mg_{1-x}Al_xB_2$ phase in a multi-phase crystal of $Mg_{0.815}Al_{0.185}B_2$ (AN210/5, see also Tabs. II - IV), determined with Image Plate data. What is striking, the $c$–axis lattice parameters calculated with two different methods for the same composition determined with EDX are located close to each other. Furthermore, they show linear dependence as a function of Al content for both single-phase crystals of $Mg_{1-x}Al_xB_2$ (0 < x < 0.1) and for crystals with precipitation of Al rich phase ($MgAlB_4$), although one can expect, that precipitation of Al rich phase introduces an error of Al estimation. Obviously the amount of the second phase in the crystals is small; therefore the error of $x$ estimation is small, too. The dependence of the $a$ lattice parameter on Al content is less systematic due to the fact, that the reflection peaks in the x-ray spectrum of both phases of $Mg_{1-x}Al_xB_2$ and of Al rich $MgAlB_4$ in the $a$ direction are placed closer, than in the $c$ direction and therefore, it is difficult to separate them. The superconducting transition temperature depends systematically on the $c$-axis lattice parameter (Fig. 4), and shows a tendency to cluster around several the $c$-axis values.

Detailed structure analysis was performed for three $Mg_{1-x}Al_xB_2$ single crystals with the following aluminum content: $x$ = 0.022 (AN229/1), $x$ = 0.044 (AN217/7), and $x$ = 0.185 (AN210/5) prepared without applying slow cooling procedure. A fourth sample with $x = 0.085$ (AN262/2), grown with slow cooling, was studied as well. Most of the x-ray measurements were carried out on a Mar-300 Image Plate system with molybdenum rotating anode at the Laboratory of Crystallography, Department of Materials, ETH Zurich. The measured images were investigated with the CrysAlis Software Package (peak search, cell finding, reconstruction of reciprocal space layers, extraction of line profiles, extraction of intensities) [22]. The raw data were corrected for Lorentz and polarization effects, no absorption corrections were applied. To check the CrysAlis results we have also used the XDS program package [23] for determination of both diffractometer and lattice parameters. Additional measurements with a CAD-4 diffractometer with a graphite monochromator and a molybdenum tube were carried out as well. SHELX97 [24] refinements of the structure were calculated using the WinGX [25] program interface.

For the two crystals AN217/7 and AN210/5 we found superstructure reflections along the $c^*$ direction in (0kl) projections of the reciprocal space. At the same time a splitting of the main reflections along $c^*$ is visible in this projections. The splitting can be interpreted as microscopic phase separation [26] into at least two phases, and it is also visible in other directions of the crystals as it is shown in Fig. 5. The crystal growth process of AN210/5, with disc made of Al placed in the crucible, obviously caused large gradient of Al content in the melt leading to large amount of the second phase shown in Fig. 6d. No additional reflections or splitting was found for the samples AN229/1 and AN262/2.

Line profiles of image plate data have been studied to characterize and visualize the phase separation of different phases along $c^*$ (shown in Fig. 6). The CrysAlis package was used to extract these line profiles out of the reconstructed (0kl) layers of the reciprocal space. The crystals AN229/1 and AN262/2 can be described with one crystal lattice and we can assume that these samples are single crystals without microscopic phase separation.

The refinement of AN229/1 and AN262/2 CAD-4 data with the $MgB_2$ structure (without additional Al positions) [27] including anisotropic atomic displacement parameters (ADP), together with refinement of the B and Mg occupancy yield acceptable R-factors (details of the refinements are given in Table II). The additional refinement of extinction parameters or the using of additional Al positions did not improve the residual factors and was therefore not used in the final refinement. For the sample AN210/5 it was possible to extract from the determined reflection list of the image plate data three partial reflection sets corresponding to the three different phases, which are visible in the (0kl) projections of the reciprocal space. A comparison of the different lattice constants that are calculated from the reflection data is given in Table III. "Phase I" corresponds to $Mg_{1-x}Al_xB_2$, "phase II" to $MgAlB_4$ and "phase III" to an impurity phase (most likely $B_2O$). As expected (taking into account the intensity ratios of "phase I" and "phase II") we obtain slightly higher R-values and more residual electron density for the refinement of AN217/7. The refinement of AN210/5 CAD-4 data yields to unreasonable R values and lattice constants due to the existence of two additional phases in this sample.

In order to separate the "phase I" and "phase II" reflections in the sample AN210/5 with the highest Al content, we have used the image plate data to obtain two intensity data sets for the crystal structure refinement of "phase I" and "phase II" (each separated with CrysAlis and than refined with SHELX). The $MgB_2$ structure model (without additional Al positions) was used to refine the ("phase I") intensity set. As one can see in Table IV, the refinement of $Mg_{1-x}Al_xB_2$ ("phase I") converged to lower R-factors ($R_1 = 0.0357$) than for the CAD-4 refinement with has taken into account only one phase with an averaged lattice (Table II); for "phase I" we obtain reasonably lattice constants and bond lengths.

The crystal structure of "phase II" can be described with the $MgAlB_4$ hexagonal superstructure, accompanied by the doubling of the $c$ axis of the $MgB_2$ structure. This $MgB_2$ superstructure was found recently with high resolution synchrotron x-ray powder diffraction measurements [28]. The refined model includes anisotropic atomic displacement parameters (ADP), together with refinement of the B, Mg and Al occupancy with yields to acceptable R-factors (details of the refinements are given in Table IV). The refinement showed Mg and Al deficiency (about 28 %) for "phase II" with yield to the chemical formula: $Mg_{0.72}Al_{0.72}B_4$ for this phase.

### B. Precipitation of a second phase

In the studies of polycrystalline $Mg_{1-x}Al_xB_2$ the solid solution has been found to phase separate for $0.1 < x < 0.4$ [13]. The tendency of a second phase to precipitate renders the growth of single-phase $Mg_{1-x}Al_xB_2$ crystals very difficult. From our crystallographic study of crystals (discussed above) we know that this is the case and the second phase of composition $MgAlB_4$ segregates as a precipitation along the $c$-axis of the crystal. In the structurally investigated crystals magnetic susceptibility measurements show only one superconducting onset indicating the existence of only one superconducting phase since, $MgAlB_4$ phase is not superconducting. HRTEM investigations indicate on precipitations of a second phase in the form of Al rich domains as well (Fig. 7). The size and shape of these domains can vary and from x-ray investigation one can conclude that up to 50% of the sample can be $MgAlB_4$ in the extreme case, when metallic disc has been used during the crystal growth as a source of Al. The amount of $MgAlB_4$ phase increases with Al content in the precursor, but differently from the observation made on polycrystalline samples even crystals with $x < 0.1$ can contain precipitation of the second phase so, we can conclude that, co-existence of one phase with variable $x$ and second one with $x = 1/2$ is typical for $Mg_{1-x}Al_xB_2$ single crystals. However, the additional phase for $0 < x < 0.1$ can be eliminate by careful tuning of the crystal growth conditions. Independently on the appearance or disappearance of an additional Al rich phase in $Mg_{1-x}Al_xB_2$, we found that for all investigated samples, simultaneously with increasing of the Al content in the precursor the Al content in the superconducting phase increases leading to decrease of $T_c$ and the $c$-axis lattice constant.

### C. High Resolution Transmission Electron Microscopy measurements

A piece of single crystal obtained in the process of AN210, in which single crystals studied with x-ray diffraction indicating on the presence of two phases, was investigated by conventional transmission electron microscopy (TEM), selected area electron diffraction (SAED) and Z-contrast (atomic mass contrast) imaging. The crystal was crushed prior to investigation. In most of the particles a large amount of defects was observed. The dislocation lines were all aligned parallel to each other in the $ab$ planes of the crystals (Fig. 8). It is not yet known whether they were formed due to the growth procedure in general or if they were caused by an accumulation of aluminum. SAED patterns even taken from highly dislocated areas did not contain additional reflections, as it would be expected if a second crystal phase exist (inset of Fig. 8). It indicates on the absence of a second crystal phase and/or of an ordering phenomenon in the highly disordered areas of studied sample. Z-contrast imaging was finally used to strengthen the image contrast caused by compositional deviations. For this technique a highly focused beam is scanned over a sample area. At the backside of the specimen those electrons, which have been scattered under a high angle, are detected with a ring detector. The intensity measured for a certain scanning position is linearly dependent on the specimen thickness and shows nearly a square dependence on the Z-value of the atomic species present in this area. Different from conventional TEM-images a Z-contrast image is bright if heavy atoms are present or if the specimen is thick. With this technique, bright, rectangular shaped areas were observed in some particles (white circles in image of Fig. 7a) while they were hardly visible on a conventional TEM image. Their longer edge was parallel to the $c$, $a$ or $b$ crystallographic axis. According to the intensity nearly following a square dependence on the Z-value of the respective atomic species these brighter areas are interpreted as areas with increased aluminum content. SAED-patterns of areas containing such rectangular shaped regions revealed the presence of superstructure reflections. Depending on the area chosen for the respective SAED-pattern, the additional diffraction intensities were observed at different positions of the reciprocal space (gray circles in Fig. 7b and 7c). While Zandbergen et al. [13] observed additional intensities in the form of a ring around the reciprocal c-axis near the positions (h,k,l+1/2), this does not appear to be the case in the present study. No systematic changes of the splitting distance of the superstructure reflections were observed with increasing distance from the undiffracted beam. Furthermore, superstructure reflections are often not only observed near positions (h,k,l+1/2), but they also accompany the main reflections (h,k,l). Future studies will have to specify, what types of ordering are present in these crystals. Astonishingly, the two phenomena, dislocations and compositional variations, appear to be independent. The dislocations (arrows in Fig. 7) are crossing the areas with different composition without any significant change of the direction of the dislocation line, additionally they are not necessarily observed along the edges of these areas and last but not least the dislocations are always aligned in the *ab* plane while the rectangular shaped areas can be aligned with the long edge also along the *c*-axis.

### D. Optimizing growth parameters

We have noticed in many cases a relatively sharp onset of diamagnetism, followed at lower temperatures by further gradual and smooth increase of the diamagnetic signal for the zero-field-cooled (ZFC) state (Fig. 9). This effect resembles weak-links effect observed earlier on Y247 samples [29]. After careful inspection of crystals with a microscope we noticed the crystals showing these effects contain very fine cracks partially crossing the crystals (Fig. 10a). One can expect, that at their narrowest part the cracks can act as weak links when the associated gap is of appropriate size. Thus the shielding current in the crystal flows through narrow cracks acting as Josephson junctions. With decreasing temperature larger parts of these cracks allow Josephson current to flow and the diamagnetic signal to grow. To test this we performed magnetization measurements on an as grown crystal, which was subsequently broken into small pieces by gentle pressing (Fig. 10b). For these pieces, the magnetization does not show any gradual change below the transition, suggesting the weak links had been removed (Fig. 9). The magnitude of the ZFC diamagnetic signal, however, is smaller. The main reason for cracks to appear in the crystals substituted with Al is the strain due to solidification of the melt. During the growth of pure $MgB_2$ crystals the strain is reduced by BN crystals forming together with $MgB_2$. Soft BN crystals separate brittle $MgB_2$ crystals and can absorb the strain caused by cooling and pressure releasing. Fewer BN crystals form when Al is present and for $x = 0.3$ almost no BN crystals can be observed. This also means, that the crystal growth mechanism changes and $Mg_{1-x}Al_xB_2$ crystals with $x > 0.3$ grow from solution in Mg-Al melt rather than following reaction (1). Without soft BN crystals, $Mg_{1-x}Al_xB_2$ crystals stick together and form the solidified Mg-Al alloy. Cooling this assembly and releasing pressure strain leads to crack formation. In order to prevent the formation of cracks cooling rate has been slowed. Additionally the temperature was kept above the melting point of the Mg-Al alloy at 800 °C, then the pressure was released, and the temperature decreased down to ambient one.

### E. Critical temperature dependence on Al content

The over-all Al content has been determined by EDX. These measurements required good crystal surfaces without residual melt or other phases covering the surface. Typically five measurements have been made on each crystal in different locations. In most cases the differences do not exceed 10% of the measured Al content. Figure 11 shows the $T_c$ dependence on Al content measured by EDX. Two growth temperatures 1860 and 1960 °C have been used, indicated by closed and open symbols, respectively. The samples for $x \leq 0.1$ are single-phase crystals, while crystals for $x > 0.1$ contain precipitation of the nonsuperconducting $MgAlB_4$ phase. The results form a relatively wide band, suggesting uncertainties in the determination of the Al content of the superconducting phase probed in the magnetization measurements. Depending of the abundance of the $MgAlB_4$ phase (for $x > 0.1$), the total amount of Al measured by EDX can vary quite much. One can notice an apparent lack of crystals with Al content between $x = 0.11$ and 0.16. It is not yet clear if the crystals of this Al content can be grown.

In the published results on polycrystalline samples the authors equate $x$ with the nominal amount of Al in the precursor, which, given the metallurgical complexities uncovered in the present study, may lead to large error in the determination of real Al content. The experimental points in Fig. 11 form a band pointing for $x \rightarrow 0$ towards $T_c = 36.5$ K and not $T_c = 38.5$ K as can be expected for pure $MgB_2$ crystals. This means, that even small, $x = 0.01$ Al content, leads to decrease of $T_c$ to 36 K. This is different from the published results on polycrystalline samples where a continuous decrease of $T_c$ as a function of Al content has been found. The reduction of $T_c$ by Al doping can by caused by two factors: (i) increase of electron concentration due to $Al^{+3}$ substituting for $Mg^{+2}$ and (ii) creation of defects by Al atoms. Even very small Al concentration seems to introduce defects in the structure. The superconducting transition broadens with increasing Al content (Fig. 12). For $T_c < 28$ K, corresponding to $x > 0.11$, $\Delta T_c$ further increases significantly, which is most likely associated with the inhomogeneous Al distribution in the scale larger than the coherence length for the investigated sample. It indicates on the dominant phase separation scenario and not on the dominant increasing amount of defects contribution scenario due to precipitation of the $MgAlB_4$ phase in the matrix of $Mg_{1-x}Al_xB_2$.

Electrical transport investigations on crystals with Al content $x > 0.1$ presented below also indicate on a tendency to multiphase formation, manifesting itself in various ways. The sharp onset of superconductivity is followed by a broad shoulder until the resistivity reaches zero at lower temperature. This effect is most probably caused by non-superconducting $MgAlB_4$ domains in the $Mg_{1-x}Al_xB_2$ structure.

**F. Electrical resistance and critical fields**

The electrical resistance of some of the Al substituted crystals is shown in Fig. 13. The data for aluminum free single crystal are presented for comparison as well. Over-all the resistivity increases when Al is incorporated, from a room temperature value of ~ 6-8 · $10^{-6}$ Ohm-cm in the unsubstituted samples to ~ 15 – 20 · $10^{-6}$ Ohm·cm when Al substitution has suppressed $T_c$ to near 30 K. The geometry factor is somewhat uncertain due to the finite dimension of the potential contacts in the 4 probe geometry and the non-uniform geometry of the crystals. In Fig. 13 the data are normalized to the value at 300 K. Starting with the unsubstituted crystal with $T_c$ = 38.6 K and a resistance ratio $rr$ = R(40K)/R(300K) of 0.084, Al substitution causes the low temperatures resistance to increase and $T_c$ to decrease. For ~ 10% Al the transition temperature is in the range of 30-32 K, and the residual resistance ratio has increased more than four times to ~ 0.35-0.4. The resistive transitions are sharp ($\leq 0.2$ K) in single-phase samples. In multi-phase samples, however, the transition reflects the inhomogeneity, not only in the usual form of a broadening but either as multiple sharp steps or as a broad "tail" extending to temperatures well below the sharp onset. While $T_c$ is not much affected by an initial increase of the resistance, we find a number of samples with $rr$ clustered around ~ 0.38-0.4 having a $T_c$ of 31-33 K, somewhat lower than extrapolated from the data points at lower $rr$. It would be interesting to investigate in detail the defect structure of these crystals, as they are close in composition to the solubility limit of Al in the present method of crystal growth.

The upper critical field $H_{c2}$ has been determined by either resistance and/or magnetization measurements (Fig. 14). When the resistive transitions are observed to be sharp, they are also found to be in good agreement with the $H_{c2}$ deduced from the sharp change in the slope of the $M(T)$ curves. In several circumstances (relative alignment of crystal, current flow and external field direction) the resistive transition is broadened due to vortex flow or surface superconductivity [30]. Details will be reported separately. In Fig. 15 an example is given of the $M(T)$ results for an Al substituted crystal with the field aligned parallel and perpendicular to the $ab$ planes. The shift in $T$ of the beginning of the diamagnetic signal is obviously dependent on the field orientation. Extensive sets of data of this type are analyzed to construct the $H_{c2} - T$ phase diagram. For two samples with different Al content the upper critical field is shown in Fig. 14. For the sample with the higher transition temperature $H_{c2}$ derived from resistance measurements is also indicated by crosses, yielding excellent agreement with the magnetically determined values of $H_{c2}$. The width of the resistive transition is indicated for a few fields by the vertical bars next to the crosses. (Here the width is conservatively defined as the temperature where the extrapolation of the $R(T)$ curve intersects with the extrapolated normal state $R(T)$ and the zero line.) Two regions of the upper critical field curve are of particular interest, as they provide complementary information about the underlying electronic scattering. The slope near $T_c$ is given by the maximum of the electrons' diffusivity in the two bands, while the limiting value of the upper critical field is determined by the minimum diffusivity. (Ref. 11) The diffusivities are different in the two bands, and they are anisotropic due the 2D and 3D nature of the bands. The slopes near $T_c$ vary in a systematic way as the Al content increases, however the magnitude of the change is rather small. For $H$ perpendicular to the $ab$-plane, the slope slightly increases from -0.1 T/K from the unsubstituted samples to ~ -0.12 T/K when $T_c$ is lowered to ~ 30 K. In contrast, the upper critical field slope decreases upon Al substitution when the field is parallel to $ab$, from ~ -0.22 T/K to ~ -0.19 T/K. Thus the anisotropy near $T_c$ slightly decreases when Al is incorporated. This is shown in the insert of Fig. 14, where the anisotropy of $H_{c2}$ for two Al substituted crystals merges at $T_c$. We have measured $H_{c2}$ resistively for several more crystals with $H$ perpendicular to the $ab$ plane. The overall trend follows the one described here, with small deviations from the generally observed trend of a decreasing d$H_{c2}$/d$T$: We would like to point out that this "reduction of anisotropy" is in stark contrast to what is observed in carbon substituted crystals, where the $H_{c2}$ slope for both orientations of the applied field rapidly increases upon C substitution, also resulting in a reduction of the "anisotropy" [15, 18]. It is rather obvious, thus, that Al substitution affects the scattering in the $H_{c2}$ determining band rather little. This is further seen in the extrapolation of $H_{c2}$ towards zero temperature. Even as the exact values of $H_{c2}$ are somewhat difficult to determine in $T$ sweeps, we observe that $H_{c2}$ perpendicular to the $ab$-plane extrapolated to zero temperature is rather close to the value in the unsubstituted samples (~ 3.5 T), suggesting the relevant minimum charge diffusivity to be unaffected by Al substitution. For $H$ parallel to the $ab$ planes $H_{c2}(0)$ is reduced considerably. Some of these observations are qualitatively similar to previous reports [15, 30]. The detailed discussion of the normal state resistivity, the slope of $H_{c2}$ and the anisotropy at $T_c$ and at lower $T$, and $H_{c2}(0)$ is beyond the scope of this report and will be given separately.

**G. Point-contact spectroscopy on Mg$_{1-x}$Al$_x$B$_2$ single crystals**

We performed systematic point-contact measurements of the energy gaps in pure MgB$_2$ crystals as well as in Mg$_{1-x}$Al$_x$B$_2$ single crystals with $0.02 \leq x \leq 0.21$. We used a modified version of the standard point-contact technique (PCS), i.e., we made the contacts by placing a small ($\varnothing \approx$ 50 μm) drop of Ag conductive paint on the crystals surface, instead of pressing a sharp metallic tip against it. This pressure-less technique (elsewhere denoted by "soft" PCS) avoids breaks in the brittle and thin crystals and ensures greater stability of the contact on thermal cycling. In all the cases described below, we made the contacts on the top crystal surface (thus injecting the probing current mainly parallel to the $c$ axis) and measured their conductance as a function of the applied voltage. Then, we normalized each curve dividing it by the relevant normal-state conductance.

Figure 16a shows an example of experimental normalized conductance curves measured at 4.6 K in a pure MgB$_2$ crystal ($x = 0$) in zero magnetic field ($G(0)$, circles). Due to the contact configuration (current parallel to the $c$ axis) the zero-field curve only presents sharp conductance maxima corresponding to $\Delta_\pi$ and no clear structures related to $\Delta_\sigma$ [2, 4]. However, it *cannot* be fitted by the standard Blonder-Tinkham-Klapwijk (BTK) model [31], unless one extends it to the two-band case by expressing the total normalized conductance $G$ as the sum of two, suitably weighed contributions from the σ and π bands: $G = w_\pi G_\pi + (1-w_\pi)G_\sigma$ [4, 32]. The fitting function contains 7 (almost) independent parameters: the gaps $\Delta_\sigma$ and $\Delta_\pi$, the barrier parameters $Z_\sigma$ and $Z_\pi$ (proportional to the potential barrier at the interface), the phenomenological broadening parameters $\Gamma_\sigma$ and $\Gamma_\pi$, and finally the weight $w_\pi$, that in pure MgB$_2$ ranges from 0.66 (for current along the *ab* planes) to 0.99 (for current along the *c* axis), as theoretically predicted [2] and experimentally verified [4]. The gap amplitudes $\Delta_\sigma$ and $\Delta_\pi$ given by the 7-parameter, two-band fit of the zero-field conductance $G(0)$ are affected by a rather large uncertainty, that can be greatly reduced if one can separate (and fit independently) the partial σ- and π-band contributions to the conductance. As previously shown, in pure MgB$_2$ this can be done by applying a magnetic field $B^*$ of about 1 T, that indeed suppresses the π-band contribution to the conductance, without affecting the σ-band gap [4, 32, 33]. The resulting conductance curve, $G(B^*)$ (squares in Fig. 16a) only contains the σ-band contribution, and thus admits a standard, one-band BTK fit with only 3 parameters. Thus, the difference $G_{\mathrm{diff}} = G(0)-G(B^*)$, shifted by 1, (triangles in Fig. 16a) only contains the π-band contribution and admits a three-parameter BTK fit as well. In the case of Fig. 16a, the values $\Delta_\sigma = 7.1 \pm 0.1$ meV and $\Delta_\pi = 2.80 \pm 0.05$ meV are obtained from the independent fit of $G(B^*)$ and $G_{\mathrm{diff}}$. As shown in Ref. 4, these values result in very close agreement with the prediction of the two-band model [2].

Figure 16b and 16c report the low-temperature, zero-field conductance curves (circles) of the *c*-axis contacts on crystals with Al content $x = 0.09$ and $x = 0.18$, respectively. Even at a first glance, it is clear that the small gap $\Delta_\pi$, related to the position of the conductance peaks, decreases on increasing $x$. At $x = 0.18$ the peaks are so close that, already at 4.2 K, the thermal smearing makes them merge in a single broad maximum, and further cooling (down to 1.88 K) was necessary to discriminate them, as shown in Fig. 16c. The zero-field curves in Fig. 16b and 16c can be perfectly fitted by the two-band BTK model (solid lines), as in pure MgB$_2$. This indicates by itself that, up to $x = 0.18$, the Mg$_{1-x}$Al$_x$B$_2$ systems is a *two-gap* superconductor. In the crystal with $x = 0.09$, one can get a more precise evaluation of the gap amplitudes by using the same procedure described above, with a field $B^* = 1$ T as in pure MgB$_2$. The fit of $G(B^*)$ (squares) and $G_{\text{diff}}+1$ (triangles) gives $\Delta_\sigma = 4.3 \pm 0.5$ meV and $\Delta_\pi = 2.30 \pm 0.15$ meV.

In the crystal with $x = 0.18$, the effect of the magnetic field looks completely different, and a field such as $B^*$ (suppressing the π-band features without affecting the σ-band gap) does not exist. For example, in the presence of a field of 1 Tesla the small-gap features do not disappear, and the change in the shape of the conductance curve seems to be rather due to a partial depression of the σ-band conductance. Fitting $G(0)$ and $G(B=1\text{T})$ with the two-band BTK model gives indeed $\Delta_\pi = 0.75 \pm 0.15$ meV for both curves, while the large gap is reduced from $\Delta_\sigma = 4.0 \pm 0.5$ meV (in zero field) to $\Delta_\sigma = 2.5 \pm 0.5$ meV (in a field of 1 T). The increase in the robustness of the π band on application of a magnetic field with respect to lower Al contents is an unexpected result and, at this stage, its origin is still not clear. Anyway it seems to indicate that, above a certain doping, radical changes involving the band structure and probably the Fermi-surface topology [34, 35] occur in our single crystals.

Figure 17a reports the temperature dependence of the conductance curves measured in a crystal with Al content $x = 0.18$. The fit of the conductance curves was performed by keeping $Z_\sigma$, $Z_\pi$, $\Gamma_\sigma$, $\Gamma_\pi$ and $w_\pi$ fixed to their low-temperature values, so that for $T > 1.88$ K the only adjustable parameters were $\Delta_\pi$ and $\Delta_\sigma$. Even with this constraint, the fit is very good up to the critical temperature of the junction, $T_c^A \approx 20$ K, at which the Andreev-reflection features disappear and the normal-state conductance is recovered. Figure 17b reports the temperature dependence of the gaps given by the fit (symbols). Within the experimental uncertainty, both gaps close at the same temperature and the shape of the $\Delta_\pi(T)$ curve indicates that the π-σ interband coupling is still rather strong [36].

Symbols in Fig. 18a represent the low-temperature gap amplitudes as a function of the bulk critical temperature $T_c$ of the crystals. Solid lines represent instead the $\Delta_\pi$ vs. $T_c$ and $\Delta_\sigma$ vs. $T_c$ curves calculated within the two-band Eliashberg theory by using the density of states at the Fermi level, $N_\sigma(E_F)$ and $N_\pi(E_F)$, and the $E_{2g}$ phonon frequency calculated as a function of the Al content from first principles [37], and adjusting the prefactor of the Coulomb pseudopotential matrix, μ, so as to reproduce the experimental $T_c(x)$ curve. It is clearly seen that the first three experimental points perfectly agree with the calculated curve, and show a progressive approaching of the two gaps that might suggest a tendency to their merging at some values of $T_c$ below 25 K [36]. However, at $x = 0.09$ ($T_c = 32.2$ K) the experimental gap values suddenly deviate from the calculated curve. This deviation is more pronounced for $\Delta_\pi$, that becomes smaller than 1 meV at the highest Al contents, $x = 0.18$ and $x = 0.21$. Such small values of $\Delta_\pi$ can be only obtained, within the two-band Eliashberg theory, by strongly reducing the intraband coupling $\lambda_{\pi\pi}$ and, consequently, the interband coupling $\lambda_{\sigma\pi}$. As a matter of fact, the dashed curves in Fig. 18a were calculated by using the same parameters ($E_{2g}$ phonon frequency, Coulomb pseudopotential prefactor, $\lambda_{\sigma\sigma}$ and $\lambda_{\pi\sigma}$) as in the case of the solid lines, but changing $\lambda_{\pi\pi}$ (and thus also $\lambda_{\sigma\pi}$) so as to fit the experimental values of the gaps. In this way, one can reproduce very well the $T_c(x)$ curve, the experimental gap behavior for $x \geq 0.09$ (with the only exception of the value of $\Delta_\sigma$ at $x = 0.09$), and also the temperature dependence of the gaps at any $x$, for example the one reported in Fig. 17b.

Figure 18b reports the $x$ dependence of the coupling constants given by first-principle calculations (solid lines) as well as those necessary to fit the experimental gap values (dashed lines). The strong decrease in $\lambda_{\pi\pi}$ above $x = 0.09$ might well be related to the observed precipitation of an Al-rich phase in the crystals. In particular, the formation of a superstructure with alternating $Mg_{1-x}Al_xB_2$ and $MgAlB_4$ layers in the crystals with $x = 0.18$ (suggested by the additional reflexes observed in x-ray diffraction patterns along the **c\*** direction, see Figs. 5e, 8b, and 8c), probably affects very little the superconducting properties in the boron planes, but could strongly weaken the inter-plane coupling. As a result, $\lambda_{\sigma\sigma}$ (that describes the coupling in the 2D σ band) is almost unchanged, while the coupling in the 3D π band, $\lambda_{\pi\pi}$, is substantially reduced. At intermediate Al contents (i.e., $x = 0.09$) the possible formation of superstructures with greater periodicity might explain the smaller decrease of the interband π-π coupling necessary to explain the point-contact results.

## IV. CONCLUSIONS

Single crystals of $MgB_2$ with Al substitution up to 28% have been grown at high pressure. X-ray refinement and HRTEM studies show a rather complicated chemistry of this substitution. By modification of the crystal growth procedure it is possible to grow single-phase crystals up to $x$ of about 0.1 - 0.12. The upper critical field $H_{c2}$ has been determined by either resistance and/or magnetization measurements. It was found that, $H_{c2}$ for the field parallel to the *ab* plane decreases with increasing Al content, while $H_{c2}$ for the field parallel to the *c* axis does not change significantly. This leads to lower $H_{c2}$ anisotropy γ. It is different from the case of carbon doping, where γ decreases with increasing carbon content, while $H_{c2}$ rises for both directions of the field.

Point-contact spectroscopy allowed the determination of the gap amplitudes, $\Delta_\pi$ and $\Delta_\sigma$, as a function of the Al content or, equivalently, of the critical temperature of the crystals. At low Al contents, the gaps perfectly agree with the predictions of the two-band Eliashberg theory, when the hardening of the $E_{2g}$ phonon mode and the changes in the densities of states of the σ and π bands due to the Al substitution are taken into account. The trend of the gaps in this low-doping region may suggest a possible gap merging at some $T_c$ below 25 K (and thus at $x > 0.21$). However, for $x \geq 0.09$ the experimental gap values strongly deviate from this trend, in a way that is highly compatible with the structural changes observed in the crystals by x-ray diffraction analyses. In particular, the decrease in the π-π intraband coupling which is necessary to explain the experimental findings might well arise from the superstructures evidenced by x-ray layer reconstructions in the crystals with the highest Al contents.

Concluding, we have given an exhaustive description of the effects of Al substitution in $MgB_2$ single crystals produced by using a high-pressure, cubic-anvil technique. We have shown how the complete set of measurements carried out with very different techniques converges towards a unified and convincing explanation of some peculiar features of these single crystals. Among these features, the segregation of a Al-rich phase – although responsible for the non-ideality of the crystals at $x \geq 0.1$ – turns out to be extremely interesting from the chemical and physical points of view, for its consequences on the electronic structure and the superconducting features of the samples. The possibility to induce this segregation in a controlled way could be a useful tool to tune the properties of this intermetallic superconductor.

**Acknowledgements**

Authors thank Dr. A. Wisniewski for fruitful discussions. This work was supported in part by the Swiss National Science Foundation, MaNEP and Polish State Committee for Scientific Research under research project for the years 2004-2006 (1 P03B 037 27). The work in Torino was done with the financial support of the INFM Project PRA "UMBRA", of the FIRB Project RBAU01FZ2P, and of the INTAS Project n. 01-617. V.A.S. acknowledges the support from RFBR and the Ministry of Science and Technology of the Russian Federation.

TABLE I. Variation of $T_c$ for $Mg_{1-x}Al_xB_2$ crystal grown with different high-pressure conditions. Each value of $T_c$ and $x$ represents the average value obtained from 5 to 10 measurements performed on different single crystals of the same process.

| Nominal Al content ($x$) in the Mg-Al melt | $T_c$ (K) for different high pressure synthesis conditions and measured Al content ($x$) | | | | | |
|---|---|---|---|---|---|---|
| | 1860° C; 0.5 h | | 1960° C; 0.5 h | | 1960° C; 0.5 h; slowly cooled to 700° C and pressure decreased to 0 | |
| | Measured Al content ($x$) | $T_c$ (K) | Measured Al content ($x$) | $T_c$ (K) | Measured Al content ($x$) | $T_c$ (K) |
| 0.05 | 0.014 | 35.9 | 0.020 | 35.6 | | |
| 0.10 | 0.028 | 35.1 | 0.035 | 34.6 | | |
| 0.15 | 0.030 | 35.0 | 0.034 | 34.8 | | |
| 0.20 | 0.048 | 34.0 | 0.076 | 33.1 | 0.078 | 32.6 |
| 0.25 | 0.078 | 32.6 | 0.092 | 31.8 | 0.114 | 29.9 |

TABLE II. Details of the CAD-4 data collection and results of the structure refinement with SHELX97 [24]. Note that for sample AN210/5 an averaged lattice (entire reflection set), without consideration of phase separation was used. This averaged lattice approach gives to unreasonable R values.

| Sample name | AN229/1 | AN217/7 | AN262/2 | AN210/5 |
|---|---|---|---|---|
| Averaged estimated chemical formula | $Mg_{0.978}Al_{0.022}B_2$ | $Mg_{0.956}Al_{0.044}B_2$ | $Mg_{0.915}Al_{0.085}B_2$ | $Mg_{0.815}Al_{0.185}B_2$ |
| Al content ($x$) | 0.022 | 0.044 | 0.085 | 0.185 |
| $T_c$ [K] | 34.7 | 34.0 | 32.4 | 25.3 |
| Crystal system | Hexagonal | | | |
| Space group | P6/mmm | | | |
| Cell constants $a$; $c$ [Å] | 3.0819(4); 3.499(2) | 3.082(3); 3.494(4) | 3.079(3); 3.488(6) | 3.062(5); 3.42(1) |
| Vol. [Å$^3$] | 28.779(18) | 28.73(4) | 28.63(7) | 27.78(12) |
| $\rho_{calc}$ [gcm$^{-3}$] | 2.650 | 2.654 | 2.664 | 2.746 |
| Crystal dimensions [mm] | 0.25 x 0.13 x 0.17 | 0.17 x 0.17 x 0.10 | 0.07 x 0.33 x 0.17 | 0.05 x 0.33 x 0.25 |
| θ range [deg] | 5.8 - 29.4 | 5.8 - 29.4 | 5.8 - 29.4 | 6.0 - 29.6 |
| $h_{min}$; $k_{min}$; $l_{min}$ | -4; 0; 0 | 0; -4; 0 | 0; -4; -4 | 0; 0; -4 |
| $h_{max}$; $k_{max}$; $l_{max}$ | 0; 3; 4 | 3; 0; 4 | 3; 0; 2 | 3; 3; 0 |
| Measured reflections | 49 | 69 | 52 | 49 |
| Number of used reflections/parameters | 30 / 7 | 30 / 7 | 30 / 7 | 30 / 7 |
| $R_1$ | 0.0309 | 0.0523 | 0.0280 | 0.0900 |
| $WR_2$ | 0.0714 | 0.1040 | 0.0662 | 0.1773 |
| GOF($F^2$) | 0.801 | 0.921 | 0.698 | 1.632 |
| Residual electron density (max; min) [e Å$^3$] | 0.392; -507 | 0.536; -1.200 | 0.276; -0.287 | 1.710; -1.303 |
| Fractional atomic coordinates, B and Mg occupation and atomic displacement parameters [Å$^3$], without additional Al positions. | | | | |
| B | x = 1/3;  y = 2/3;  z = ½ | | | |
| Occupation, B | 1.010(12) | 1.00(2) | 1.007(16) | 1.02(4) |
| Mg | x = 0;  y = 0;  z = 0 | | | |
| Occupation, Mg | 0.970(11) | 0.99(2) | 0.980(14) | 1.01(3) |
| $U_{11}$, B | 0.0054(6) | 0.0072(12) | 0.0052(8) | 0.009(2) |
| $U_{33}$, B | 0.0071(12) | 0.0031(18) | 0.0100(13) | 0.008(3) |
| $U_{12}$, B | 0.0027(3) | 0.0036(6) | 0.0026(4) | 0.0047(10) |
| $U_{11}$, Mg | 0.0048(4) | 0.0076(9) | 0.0050(5) | 0.0081(14) |
| $U_{33}$, Mg | 0.0047(7) | 0.0007(10) | 0.0079(7) | 0.0026(17) |
| $U_{12}$, Mg | 0.0024(2) | 0.0038(5) | 0.0025(3) | 0.0040(7) |
| Mg-B bond length [Å] | 2.4952(7) | 2.4934(15) | 2.490(3) | 2.459(5) |
| B-B bond length [Å] | 1.7793(2) | 1.7791(6) | 1.7778(18) | 1.7689(16) |

TABLE III. Cell constants for the three phases in sample AN210/5 calculated for three sets of the reflections. $Mg_{1-x}Al_xB_2$ ("phase I", 194 reflections), $MgAlB_4$ ("phase II", 64 reflections) and impurity ("phase III", 32 reflections).

| Phase | $a$ [Å] | $b$ [Å] | $c$ [Å] | $\alpha$ [deg] | $\beta$ [deg] | $\gamma$ [deg] | $V$ [Å$^3$] |
|---|---|---|---|---|---|---|---|
| I | 3.083(1) | 3.080(1) | 3.459(1) | 89.97(3) | 89.99(3) | 119.97(4) | 28.46 |
| II | 3.046(3) | 3.048(3) | 6.724(6) | 90.00(7) | 90.14(7) | 119.80(9) | 54.17 |
| III | 2.892(3) | 2.900(3) | 7.108(6) | 89.99(8) | 89.96(7) | 120.03(11) | 51.46 |

TABLE IV. Details of the Two-Phase refinement (SHELX97 [24]) of the sample AN210/5 ($T_c$ = 25.3 K) measured with the Mar-300 Image Plate system. Averaged estimated chemical formula: $Mg_{0.815}Al_{0.185}B_2$. Crystal dimensions [mm]: 0.05 x 0.33 x 0.25.

| Sample name | AN210/5 | |
|---|---|---|
| | $Mg_{1-x}Al_xB_2$ ("phase I") | $Mg_{0.72}Al_{0.72}B_4$ ("phase II") |
| Crystal system | Hexagonal | Hexagonal |
| Space group | P6/mmm | P6/mmm |
| Cell constants $a$; $c$ [Å] | 3.082(1); 3.459(1) | 3.047(3); 6.724(6) |
| Vol.[Å$^3$] | 28.454(15) | 54.17(9) |
| $\rho_{calc}$ [gcm$^{-3}$] | 2.680 | 2.898 |
| Range of angles $\theta$ [deg] | 5.9 - 21.5 | 3.0 - 21.8 |
| $h_{min}$; $k_{min}$; $l_{min}$ | -3; -3; -3 | -3; -3; -7 |
| $h_{max}$; $k_{max}$; $l_{max}$ | 3; 3; 3 | 3; 3; 7 |
| Measured reflections | 123 | 235 |
| Used reflections/parameters | 11 / 7 | 23 / 11 |
| $R_1$ | 0.0357 | 0.1020 |
| $wR_2$ | 0.0987 | 0.3477 |
| GOF($F^2$) | 1.483 | 4.488 |
| Residual electron density (max; min) [e/Å$^{-3}$] | 1.460; -0.890 | 1.015; -0.641 |
| Fractional atomic coordinates, occupation and atomic displacement parameters [Å$^3$] | | |
| x; y; z B | x = 1/3; y = 2/3; z = ½ | x = 1/3; y = 2/3; z = 0.2554(7) |
| Occupation, B | 1.02(5) | 0.93(8) |
| $U_{11}$; $U_{33}$; $U_{12}$ B | 0.027(8); 0.036(7); 0.013(4) | 0.015(12); 0.13(2); 0.008(6) |
| x; y; z Mg | x = 0; y = 0; z = 0 | x = 0; y = 0; z = 0 |
| Occupation, Mg | 0.97(4) | 0.717(9) |
| $U_{11}$; $U_{33}$; $U_{12}$ Mg | 0.011(4); 0.031(4); 0.0053(18) | 0.0088(12); 0.045(3); 0.0044(6) |
| x; y; z Al | without additional Al positions | x = 0; y = 0; z = 0.5 |
| Occupation, Al | | 0.719(10) |
| $U_{11}$; $U_{33}$; $U_{12}$ Al | - | 0.0087(13); 0.037(3); 0.0043(6) |
| Mg-B bond length [Å] | 2.4807(7) | 2.455(4) |
| Al-B bond length [Å] | - | 2.405(4) |
| B-B bond length [Å] | 1.7782(7) | 1.7574(17) |

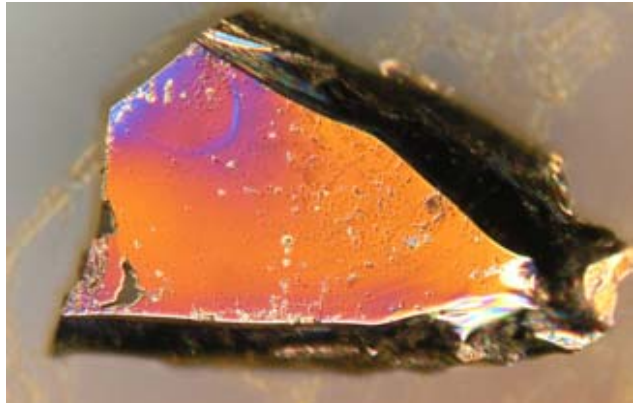

FIG. 1 (Color online). (Mg,Al)B$_2$ single crystal of about 1mm length.

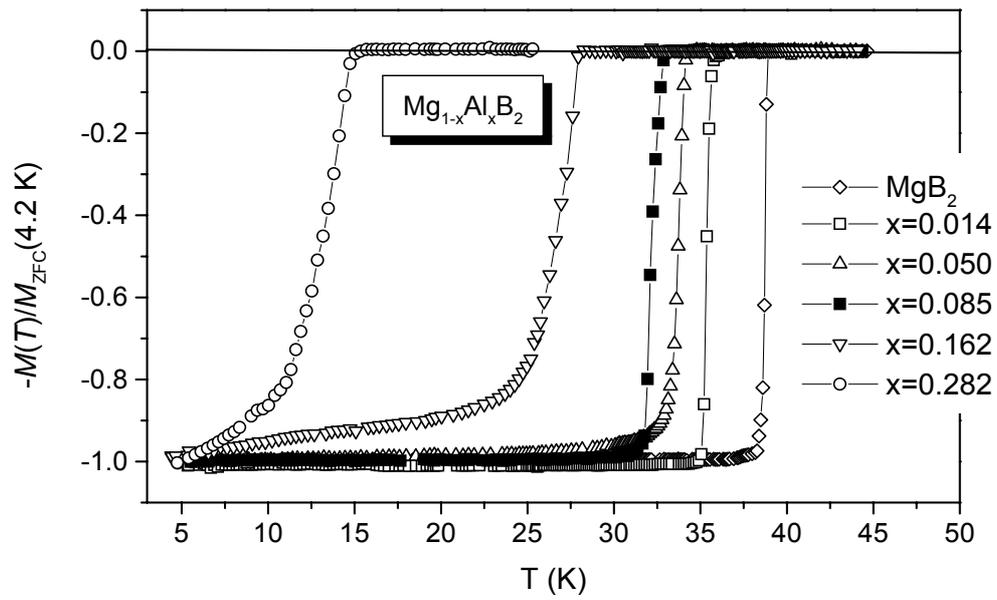

FIG. 2. Normalized diamagnetic signal of Mg$_{1-x}$Al$_x$B$_2$ crystals with various Al content. The measurements were performed at dc field of 3 Oe in zero-field-cooled (ZFC) mode, i.e., with increasing temperature after cooling sample in zero magnetic field.

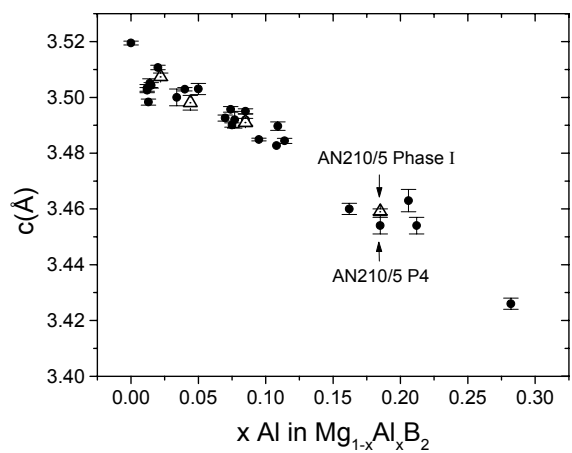
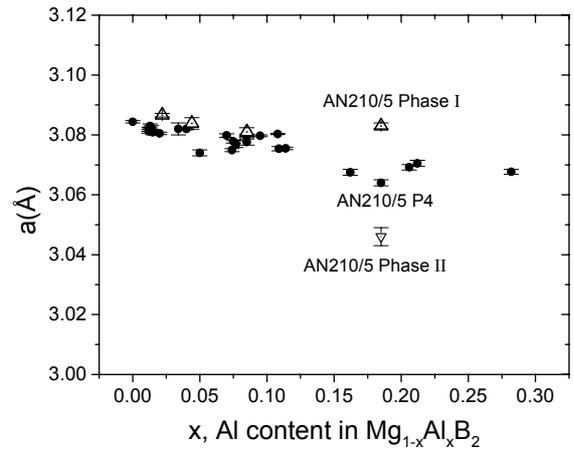

a)  b)

FIG. 3. Lattice parameter *c* (panel a) and lattice parameter *a* (panel b) both versus Al content determined with EDX. Full circles indicate lattice parameters for single-phase samples or lattice parameters of the main phase for multiphase samples (phase separation at Al contents x > 0.1; in this content range the lattice parameter are determined for an averaged lattice), determined with Siemens P4 diffractometer data. Open up triangles indicate lattice parameters for single-phase samples or lattice parameters of the main $Mg_{1-x}Al_xB_2$ phase for multiphase samples, determined with Mar-300 Image Plate data. With this method it was possible to determine for one crystal with higher Al content of *x* = 0.185 (crystal AN210/5; also measured with Siemens P4) the lattice parameter for the $MgAlB_4$ phase (lattice constant *a*; open down triangle; see also Table II).

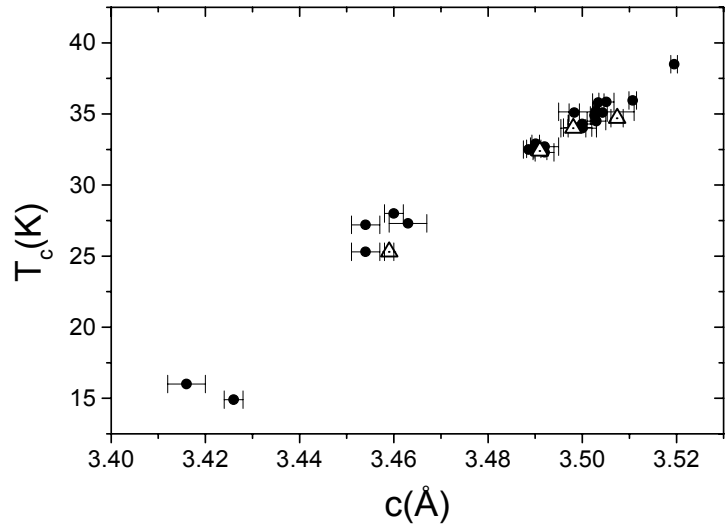

FIG. 4. Superconducting transition temperature as a function of the lattice parameter $c$ for the single-phase samples or of the lattice parameter $c$ of the main $Mg_{1-x}Al_xB_2$ phase for multiphase samples, determined with the Siemens P4 diffractometer (full circles) and with the Mar-300 Image Plate (open triangles).

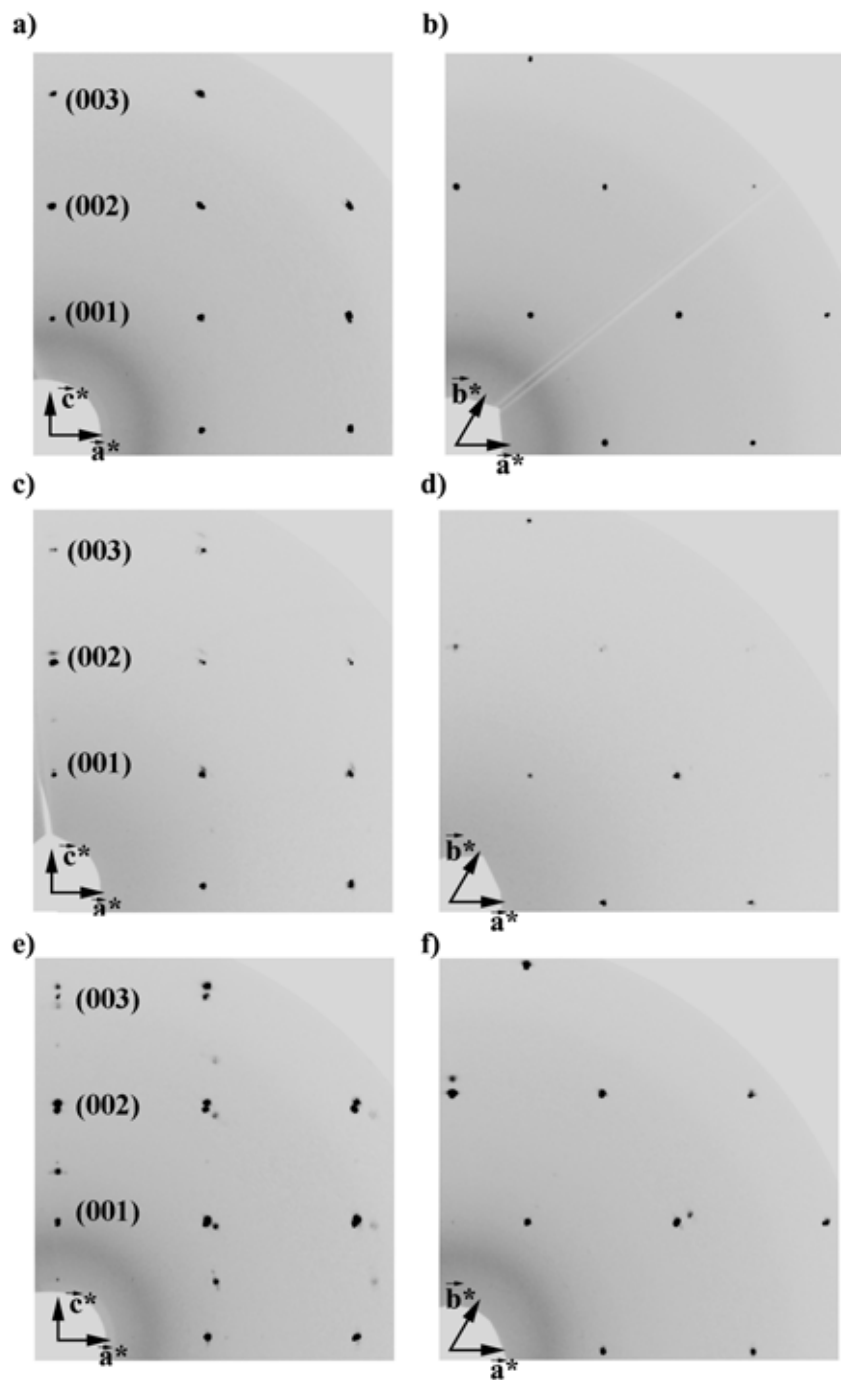

FIG. 5. Reconstructed layer (0kl) [a, c and e] and (hk0) [b, d and f] of $Mg_{1-x}Al_xB_2$ single crystals with varying aluminum content (reconstructed from 180 frames, 600 s exposure time and $\Delta\varphi = 1°$ per frame; sample to detector distance, 150 mm). Panels a and b presents results obtained for crystals with $x = 0.022$ (AN229/1), panels c and d for crystals with $x = 0.044$ (AN217/7), and panels e and f for crystals with $x = 0.185$ (AN210/5). Both crystals with high aluminum content of $x = 0.044$ (AN217/7) and $x = 0.185$ (AN210/5) show additional reflexes.

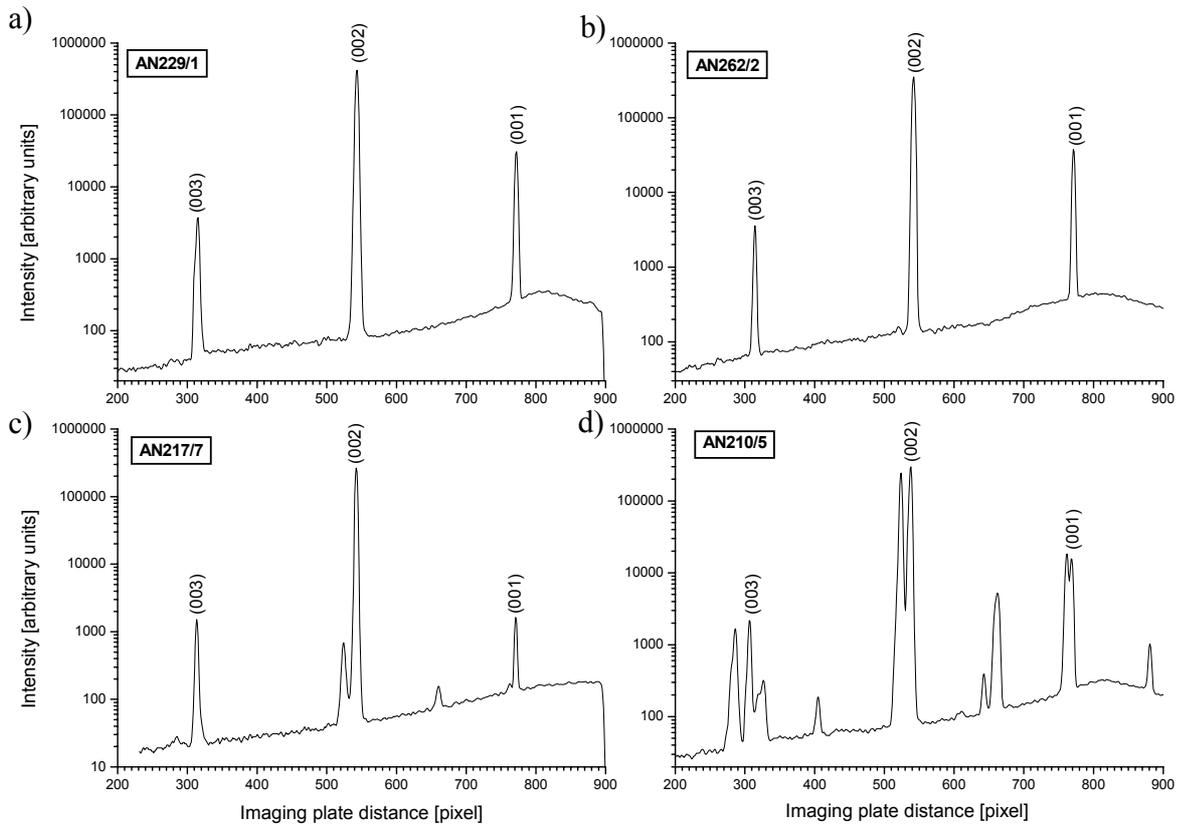

FIG. 6. Line profiles (intensity on logarithmic scale) along **c**$^*$ of reconstructed (0kl) layers of $Mg_{1-x}Al_xB_2$ single crystals with various aluminum content. Panel a presents data obtained for the crystal with $x = 0.022$ (AN229/1), panel b for the crystal with $x = 0.085$ (AN262/2), panel c for the crystal with $x = 0.044$ (AN217/7), and panel d for the crystal with $x = 0.185$ (AN210/5). The indices of the reflections (00l) of $Mg_{1-x}Al_xB_2$ ("phase I"; $c \approx 3.46$ Å), $Mg_{0.5}Al_{0.5}B_2$ ("phase II"; $c \approx 6.72$ Å) and an additional impurity ("phase III"; $c \approx 7.11$ Å) are shown. The peaks of additional phases are absent for the samples AN229/1 and AN262/2, they are slightly developed for AN217/7 and are well visible for AN210/5.

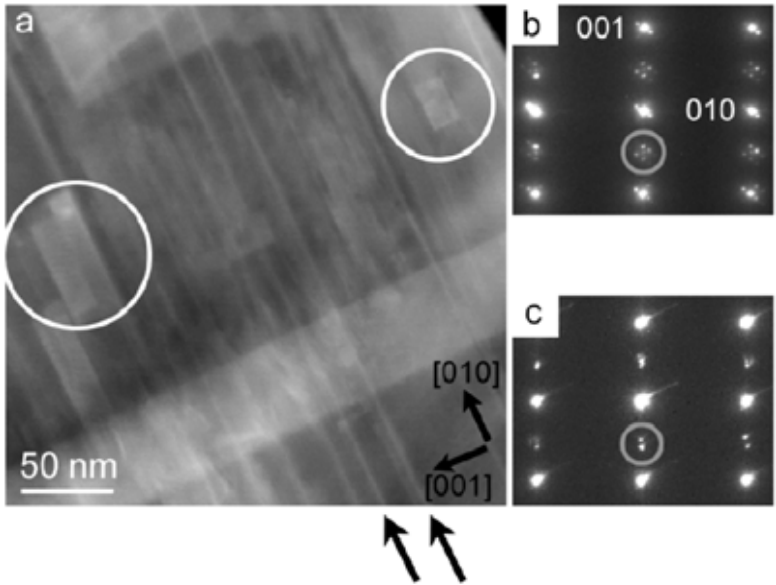

FIG. 7. Panel a: Z-contrast image of an area containing rectangular shaped areas with brighter contrast (white circles) for the crushed crystal of AN210. These areas are supposed to contain a higher Al-content. Panels b and c: Various types of superstructure reflections (gray circles) observed in diffraction patterns taken from regions containing such areas. Dislocations (arrows) present in the particle appear to be independent of the respective composition.

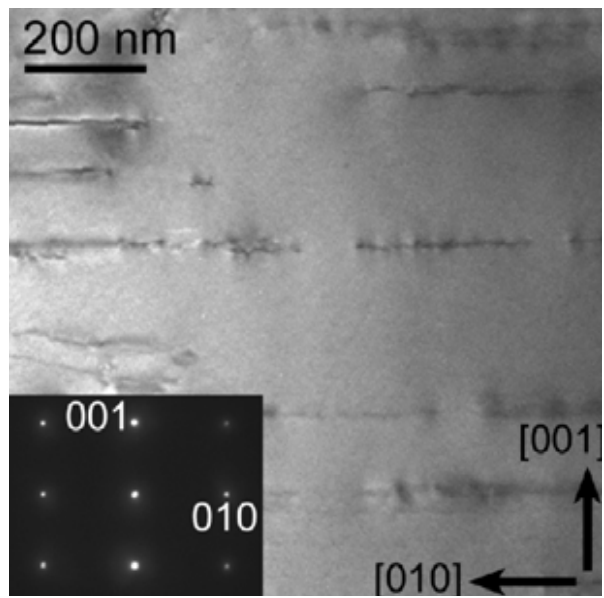

FIG. 8. Conventional TEM image of a particle containing dislocations obtained for the crushed crystal of AN210. The dislocation lines are aligned parallel to each other in the *ab*-plains. The diffraction pattern of this particle (inset) does not contain superstructure reflections.

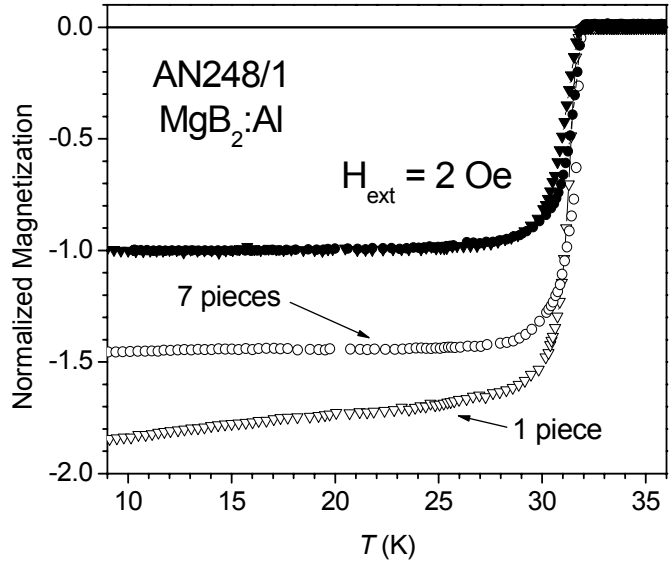

FIG. 9. Magnetization measurements performed in ZFC (open symbols) and FC (closed symbols) mode in dc field of 2 Oe on $Mg_{1-x}Al_xB_2$ single crystal in 1 piece (triangles down) and after crushing it in 7 pieces (circles). The $M(T)$ data were normalized to magnetization at 5 K measured in FC mode (Meissner signal).

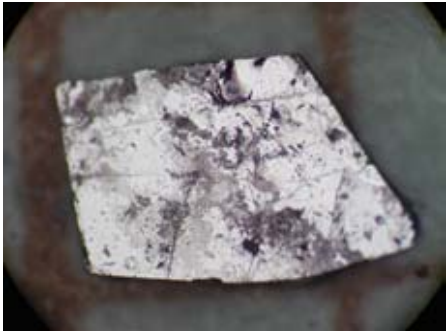

a)

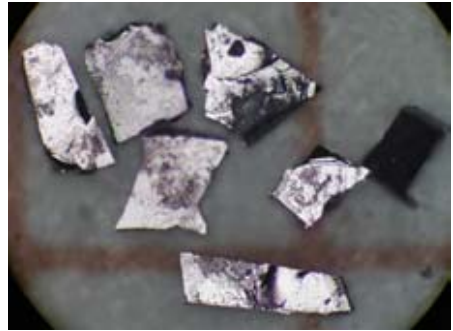

b)

FIG. 10 (Color online). Panel a: $Mg_{1-x}Al_xB_2$ crystal showing weak-link effects in magnetization measurements (Fig. 9). Panel b: the same crystal after breaking does not show weak links.

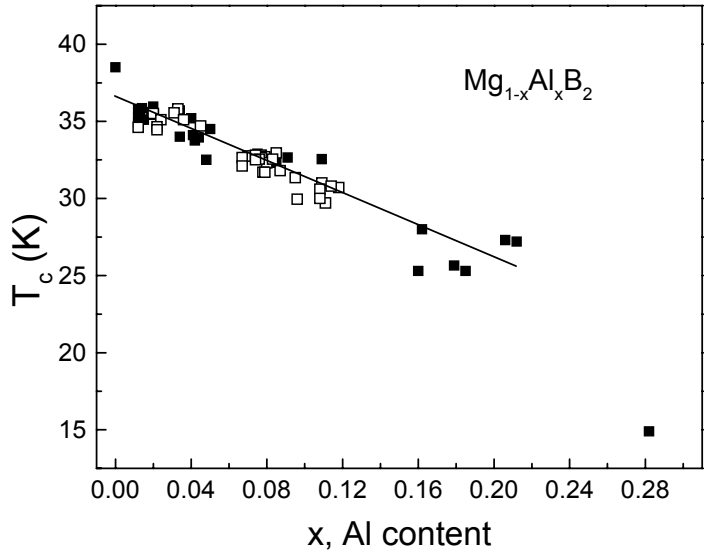

FIG. 11. $T_c$ dependence on Al content, determined by EDX, for the $Mg_{1-x}Al_xB_2$ single crystals grown at $T = 1860°C$: (closed squares) and $T = 1960°C$: (open squares).

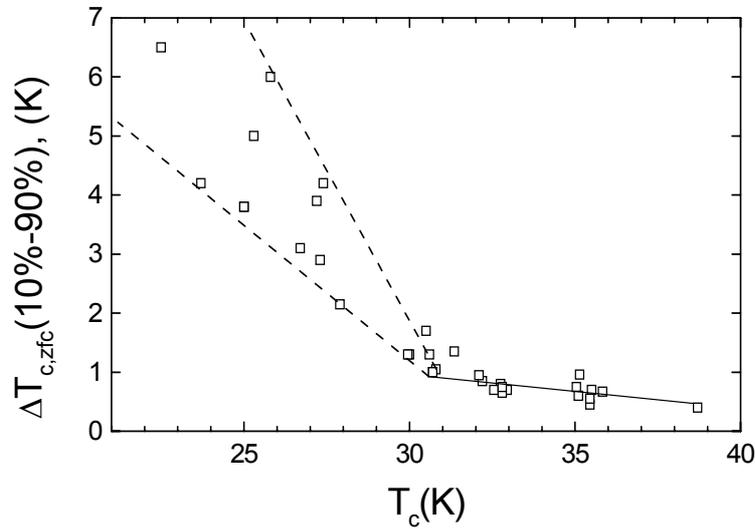

FIG. 12. Width of the superconducting transition $\Delta T_c$ as a function of $T_c$. Each point represents the average value obtained for several (3 to 10) crystals from the same batch.

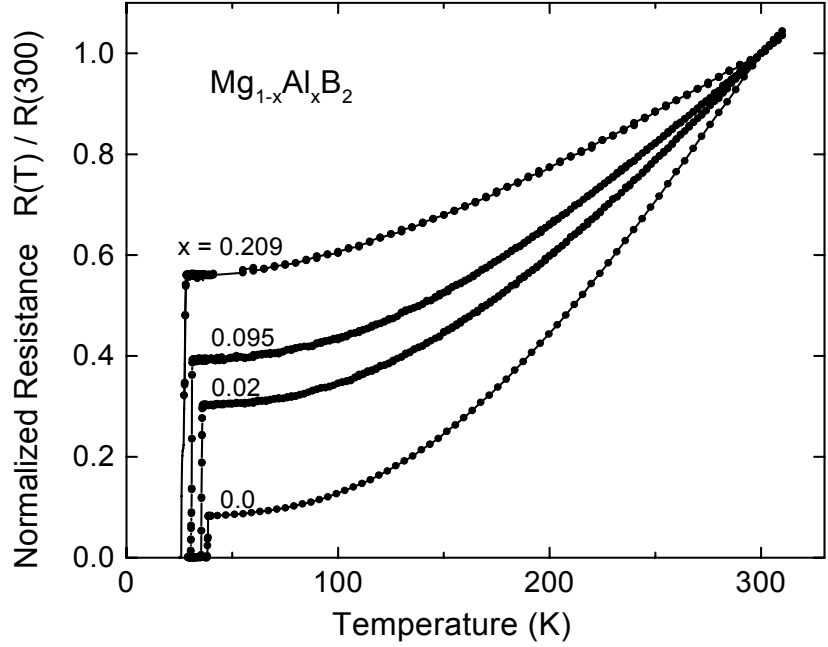

FIG. 13. Electrical resistance of $Mg_{1-x}Al_xB_2$ crystals. The room temperature resistivity increases from $\rho(300\ K) \approx 6\text{-}8 \cdot 10^{-6}$ $\Omega$cm, for nonsubstituted samples, to about $15\text{-}20 \cdot 10^{-6}$ $\Omega$cm for samples with $x \approx 0.10$ ($T_c \approx 30$ K). A pure nonsubstituted crystal has $T_c = 38.6$ K, $\rho(40K) = 0.48$ ($\pm 0.03$) $\cdot 10^{-6}$ $\Omega$cm, and a resistance ratio $rr = R(40K)/R(300K) = 0.084$.

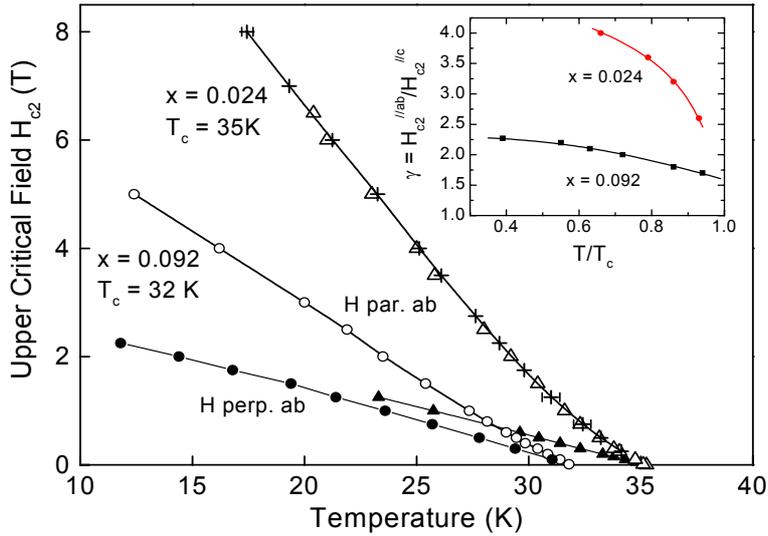

FIG. 14. Upper critical fields determined from magnetic and/or resistance (crosses) measurements for the field parallel to the $ab$–plane and the $c$ axis for two $Mg_{1-x}Al_xB_2$ crystals with different Al content. Insert shows the anisotropy of $H_{c2}$ for the same crystals.

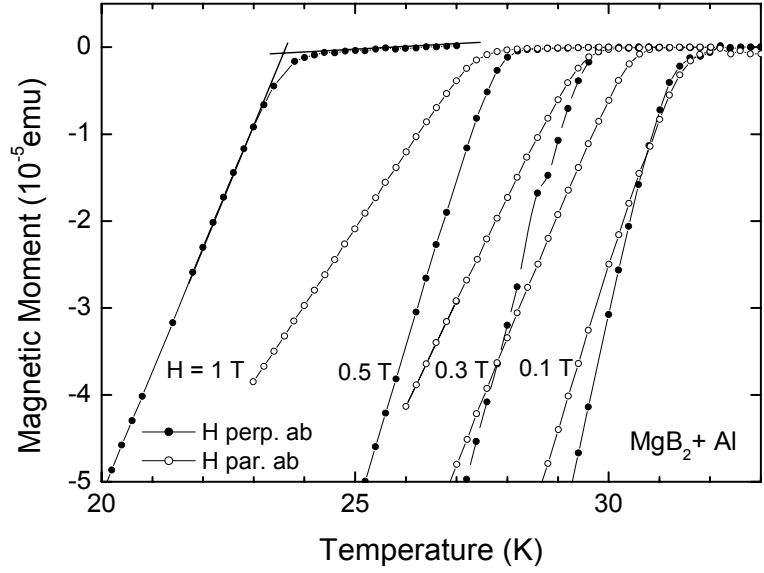

FIG. 15. Magnetic moment for Al substituted crystal with the field aligned parallel and perpendicular to the *ab* planes. Extensive sets of data of this type are analyzed to construct the *H-T* phase diagram (see Fig. 14).

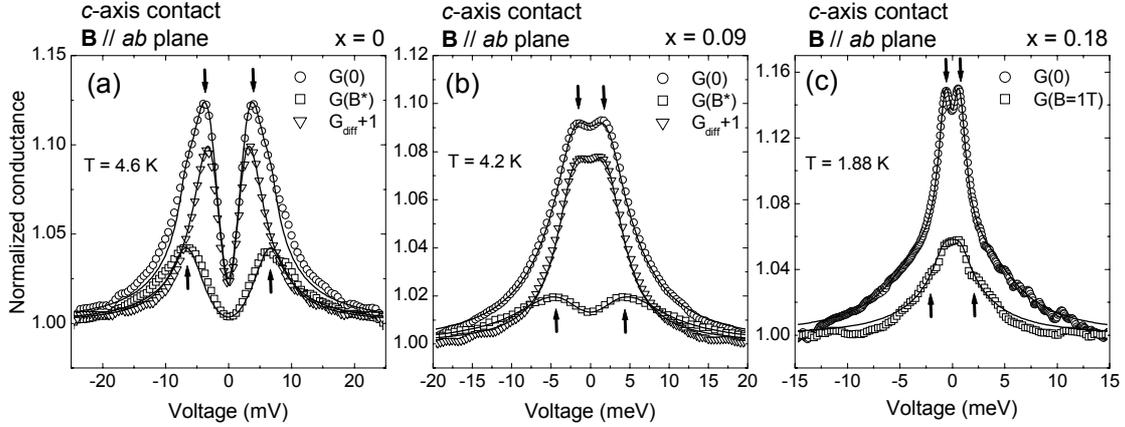

FIG. 16. Low-temperature experimental normalized conductance curves for pure $MgB_2$ (a) and Al-doped $MgB_2$ with $x = 0.09$ (b) and $x = 0.18$ (c). All panels report the conductance curves measured in zero magnetic field (circles), and in the presence of a magnetic field of 1 T parallel to the *ab* planes (squares), compared with the relevant two-band or single-band BTK fit (lines). For details see the text. In panel a and b, triangles indicate the difference between the previous two experimental curves, compared to its single-band BTK fit (lines).

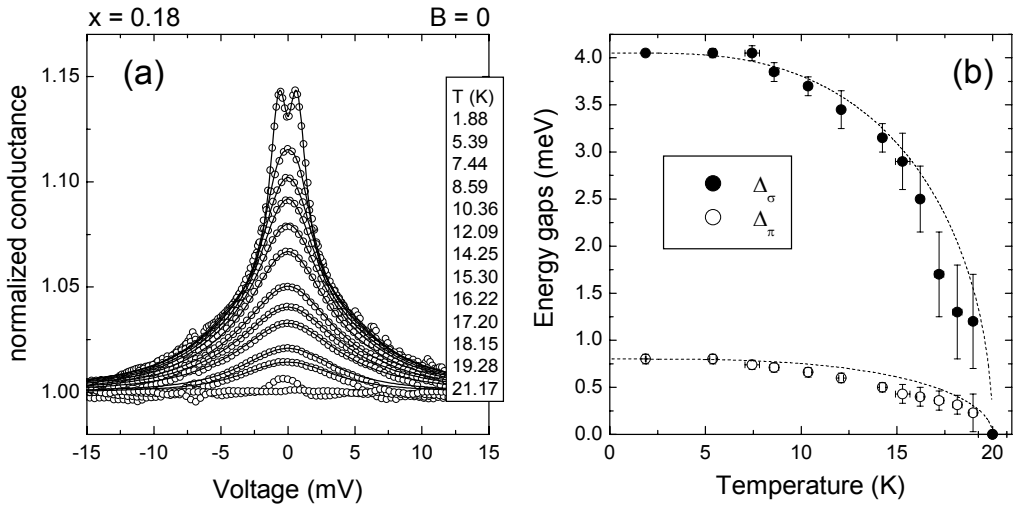

FIG. 17. Panel a: An example of temperature dependence of the normalized conductance curves of a point contact made on a Al-doped $MgB_2$ sample with $x = 0.18$. Circles represent experimental data, lines the corresponding two-band BTK best-fitting curves. Note that the curves reported here are a subset of all the curves measured. Panel b: Temperature dependence of the gaps $\Delta_\pi$ and $\Delta_\sigma$ as obtained from the fit of the curves in panel a. Lines are BCS-like $\Delta(T)$ curves reported for comparison. Note that, within the experimental uncertainty, both gaps close at the same temperature.

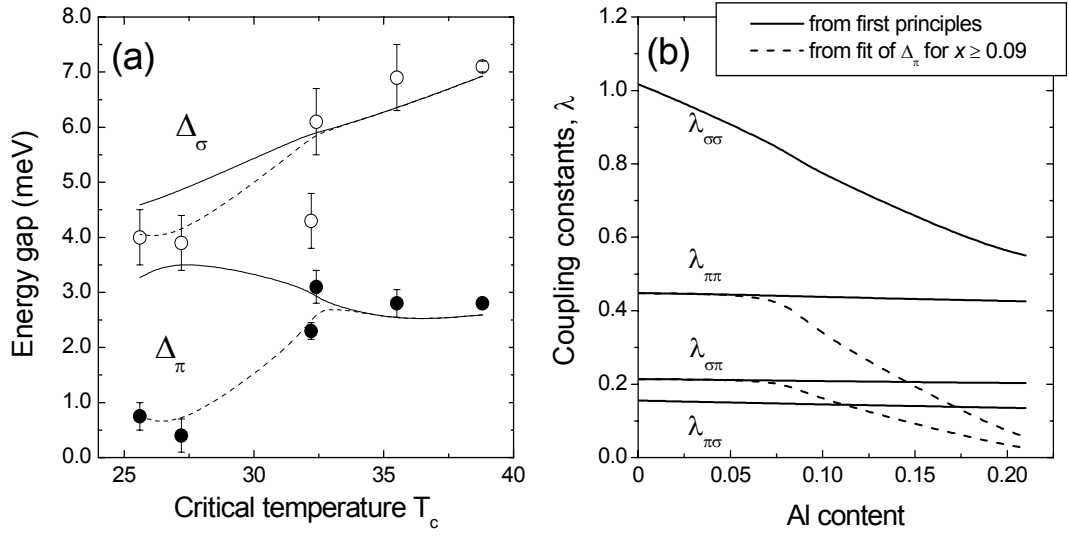

FIG. 18. Panel a: Low-temperature energy gaps measured in Al-doped single crystals as a function of the bulk critical temperature $T_c$. Solid line: values of the gaps calculated by using the frequency of the $E_{2g}$ phonon mode and the density of states calculated from first principles in Ref. 37, and adjusting the Coulomb pseudopotential prefactor so as to fit the experimental $T_c(x)$ curve. Dashed lines: best fit of the experimental gaps in the high-doping region, obtained by reducing the value of the π-π coupling constant (and consequently also the σ-π one). Panel b: Dependence of the intraband and interband coupling constants λ on the Al content. Solid and dashed lines refer to the same cases as in panel a. The strong reduction of $\lambda_{\pi\pi}$ and $\lambda_{\sigma\pi}$ from the values calculated from first principles, necessary to reproduce the experimental gap values, is clearly seen.